\documentclass[aps,prb,twocolumn,superscriptaddress,longbibliography]{revtex4-2}
\usepackage{graphicx}
\usepackage{latexsym}
\usepackage{amssymb}
\usepackage{amsmath}
\usepackage{amsfonts}
\usepackage{upgreek}
\usepackage{float}
\usepackage{ulem}
\usepackage{bm}
\usepackage{multirow}
\usepackage{color}
\usepackage{booktabs}  
\usepackage[T1]{fontenc}
\usepackage{hyperref}

\hypersetup{
colorlinks = true,
linkcolor = [rgb]{0.70,0.13,0.13},
citecolor = [rgb]{0.13,0.55,0.13},
urlcolor  = [rgb]{0.25, 0.41, 0.88}}

\begin{document}

\title{Generalized Aubry-Andr\'{e}-Harper model with power-law quasiperiodic potentials}

\author{Ya-Nan Wang}
\affiliation{College of Science, Nanjing University of Aeronautics and Astronautics, Nanjing, 211106, China}
\affiliation{Key Laboratory of Aerospace Information Materials and Physics (NUAA), MIIT, Nanjing 211106, China}

\author{Wen-Long You}
\thanks{Corresponding author: wlyou@nuaa.edu.cn}
\affiliation{College of Science, Nanjing University of Aeronautics and Astronautics, Nanjing, 211106, China}
\affiliation{Key Laboratory of Aerospace Information Materials and Physics (NUAA), MIIT, Nanjing 211106, China}

\author{Zhihao Xu}
\thanks{Corresponding author: xuzhihao@sxu.edu.cn}
\affiliation{Institute of Theoretical Physics and State Key Laboratory of Quantum Optics and Quantum Optics Devices, Shanxi University, Taiyuan, 030006, China}
\affiliation{Collaborative Innovation Center of Extreme Optics, Shanxi University, Taiyuan, 030006, China}

\author{Gaoyong Sun}
\thanks{Corresponding author: gysun@nuaa.edu.cn}
\affiliation{College of Science, Nanjing University of Aeronautics and Astronautics, Nanjing, 211106, China}
\affiliation{Key Laboratory of Aerospace Information Materials and Physics (NUAA), MIIT, Nanjing 211106, China}

\begin{abstract}
We investigate a generalized Aubry-Andr\'{e}-Harper (AAH) model  with non-reciprocal hopping and power-law quasiperiodic potentials $V(i) = V\left[ \cos(2\pi \beta i) \right]^p$. Our study reveals that the interplay between nonreciprocity, quasiperiodicity, and the power-law exponent $p$ gives rise to a variety of phase transitions and localization phenomena. In the Hermitian case, the system undergoes a direct transition from extended to localized phases for $p=1, 2$, while for \(p \geq 3\), an intermediate mixed phase emerges, characterized by the coexistence of extended and localized states and the presence of mobility edges. Importantly, we find that prominent high-IPR states associated with well-resolved spectral gaps appear at specific energy levels, whose positions are captured by the relation \(x_n = n\beta - \lfloor n\beta \rfloor\), for low-order $n$. In the non-Hermitian regime, the energy spectrum becomes complex and the \(\mathcal{PT}\) transition coincides with the extended-to-localized phase boundary for \(p = 1, 2\), whereas for \(p \geq 3\), \(\mathcal{PT}\)-symmetry breaking occurs at the mixed-to-localized phase transition. This work reveals how power-law quasiperiodic potentials and non-reciprocal hopping govern phase transitions, providing new insight into localization phenomena of quasiperiodic systems.

\end{abstract}

\maketitle

\section{Introduction}
Anderson localization is a key phenomenon observed in noninteracting disordered systems \cite{anderson1958absence,abrahams1979scaling,kramer1993localization,evers2008anderson}, in which strong disorder leads to the spatial localization of particle wavefunctions. In disordered systems with dimensionality $d\leq2$, the conductance decreases with system size and vanishes in the thermodynamic limit according to the scaling theory of localization \cite{abrahams1979scaling}. This implies that all electronic states are localized in such low-dimensional systems with $d\leq2$, whereas extended states can occur only for $d>2$. In contrast, quasiperiodic systems, lying between periodic and fully disordered systems, are distinct. For example, the one-dimensional quasiperiodic Aubry-Andr\'{e}-Harper (AAH) model \cite{aubry1980analyticity, harper1955single}, owing to its self-duality, undergoes a transition from extended to localized states.

Experimental realizations of the AAH model have been achieved in various platforms, including photonic systems \cite{lahini2009observation,kraus2012topological,verbin2013observation,gao2025probing}, ultracold atoms \cite{luschen2018single,an2018engineering} and superconducting qubits \cite{xiang2023simulating,li2023observation}.
The AAH model, renowned for its self-duality and analytical solvability, motivated numerous studies, including long-range hopping \cite{xu2021non,roy2021fraction,peng2023power,liu2024emergent,peng2025long}, generalized on-site potentials \cite{ganeshan2015nearest,qi2023multiple,banerjee2025emergence}, and the $p$-wave pairing \cite{liu2021fate,tong2021dynamics,lv2022quantum,lv2022exploring,fraxanet2022localization,gandhi2025superconducting}.
In particular, it has been revealed that generalized AAH models can exhibit an energy-dependent mobility edge \cite{ganeshan2015nearest,biddle2010predicted,wang2020one,wang2021duality,an2021interactions,liu2021exact,zhou2023exact,liu2024quantum,wang2024exact,jiang2025localization}, absent in the original AAH model, which separates extended and localized states.
These developments have significantly advanced the understanding of the Anderson transition in disordered systems.

Recently, non-Hermitian systems characterized by biorthogonal eigenstates and complex energy spectra have attracted growing interest, revealing a wealth of novel phenomena such as $\mathcal{PT}$-symmetry breaking \cite{bender1998real,guo2009observation,el2018non,wu2019observation,hu2024geometric,cai2025quantum}, exceptional points \cite{zhou2018observation,yang2019non,miri2019exceptional,xiao2021observation,wu2024third},  the non-Hermitian skin effect \cite{yao2018edge,zhang2020correspondence,li2020critical,yokomizo2019non,yang2020non,okuma2020topological,liang2022dynamic}, and spectral topology \cite{yin2018geometrical,gong2018topological,wang2021topological,wang2021generating}. The interplay between quasiperiodicity and non-Hermiticity, realized through nonreciprocal hopping or complex quasiperiodic potentials, provides a versatile platform for exploring localization and topological phenomena \cite{zeng2017anderson,zeng2020winding,zeng2020topological,jiang2019interplay,liu2020generalized,zhai2020many,liu2021localization,zhai2021cascade,zhai2022nonequilibrium,cai2022localization,dai2023emergence,li2024asymmetric,zhou2024entanglement,ren2024identifying,sun2024hybrid,wang2025family,padhi2025anomalous,liu2020non,Tomasi2022nonPR}.
In particular, in the original non-Hermitian AAH model, the Anderson and $\mathcal{PT}$ transitions are expected to coincide \cite{jiang2019interplay,zeng2025fidelity}, raising the question of whether this behavior is universal or merely coincidental. 

Motivated by these questions, we study a generalized AAH model featuring a power-law quasiperiodic potential $V(i) = V\left[ \cos(2\pi \beta i) \right]^p$. For $p=2$, we find that the system mainly modifies the oscillation frequency and shifts the critical point of the phase transition, while still exhibiting a direct transition from extended to localized states without the appearance of a mobility edge, similar to the original AAH model ($p=1$). In contrast, for $p \ge 3$, an intermediate mixed phase emerges, characterized by the coexistence of extended and localized states and the presence of mobility edges.
Moreover, we observe that when the system size is not a Fibonacci number, the single-particle inverse participation ratio (IPR) exhibits pronounced peaks at specific energy levels. The positions of these high-IPR states are governed by a universal relation $x_n = n\beta - \lfloor n\beta \rfloor$ which holds for all exponents $p$ and provides an accurate prediction of the mobility edges. 

For the non-Hermitian AAH model with a power-law quasiperiodic potential, we find that the transition from the extended to localized phases coincides with the $\mathcal{PT}$ transition for $p=1, 2$. However, for $p \ge 3$, the system develops an intermediate mixed phase in which extended and localized states coexist, and the $\mathcal{PT}$ symmetry-breaking transition no longer coincides with the extended-to-localized transition, but instead occurs between the mixed and localized phases. The emergence of the mixed phases and the corresponding transition are captured by the density of states with complex energies. These findings deepen our understanding of the non-Hermitian AAH model and provide new insights into localization phenomena in quasiperiodic systems.

The article is organized as follows. In Sec.~\ref{sec:model}, we present the non-Hermitian AAH model with a power-law quasiperiodic potential. In Sec.~\ref{sec:methods}, we define the quantities used in the numerical simulations, including the IPR, fractal dimension, and density of states. In Sec.~\ref{sec:results}, we investigate the results for both the Hermitian and non-Hermitian AAH models. In Sec.~\ref{sec:wave}, we discuss the wave-packet spreading. Finally, in Sec.~\ref{sec:Con}, we provide a summary of the study.

\section {Model}
\label{sec:model}
We investigate a one-dimensional AAH model with nonreciprocal hopping and a power-law quasiperiodic on-site potential. The Hamiltonian of the system is defined as
\begin{equation}
\label{eq:Ham}
    H=-t\sum_{i}(e^{-g}c_i^{\dagger}c_{i+1}+e^{g}c_{i+1}^{\dagger}c_{i})+\sum_{i} V(i) n_{i},
\end{equation}
where $c_{i}^{\dagger}$ and $c_{i}$ are the creation and annihilation operators of a single fermionic particle, respectively, and $n_i=c_{i}^{\dagger}c_{i}$ represents the particle number operator.
$te^{-g}$ and $te^{g}$ denote the nonreciprocal nearest-neighbor hopping amplitudes.
The on-site potential is given by $V(i)=V\left[ \cos(2\pi \beta i) \right]^p$, where $V$ denotes the strength of the quasiperiodic potential and $\beta$ is an irrational number.
$p$ denotes the exponent of the on-site potential. It is precisely the choice of an irrational number that enables the formation of a quasiperiodic potential. Using the Fibonacci sequence defined by $F_{u} = kF_{u-1} + F_{u-2}$, with initial conditions \( F_0 = 0 \) and \( F_1 = 1 \), the irrational number \( \beta \) is given by the limit \( \beta = \lim\limits_{u \to \infty} F_{u-1} / F_{u} \). This yields the golden ratio \( \beta_g = (\sqrt{5} - 1)/2 \) for \( k=1 \), the silver ratio \( \beta_s = \sqrt{2} - 1 \) for \( k=2 \), and the bronze ratio \( \beta_b = (\sqrt{13} - 3)/2 \) for \( k=3 \).
Throughout the paper, we take \( t = 1 \) and use periodic boundary conditions in the numerical simulations.

For \( g = 0 \) and \( p = 1 \), the Hamiltonian (\ref{eq:Ham}) reduces to the standard AAH model. The system undergoes a phase transition at the critical point \( V = 2t \)~\cite{jiang2019interplay} owing to its self-duality \cite{aubry1980analyticity, harper1955single}. Specifically, for \( V < 2t \), all states are extended, whereas for \( V > 2t \), all states are localized. When $g \neq 0$, the model becomes a non-Hermitian AAH system. It has been shown that the eigenvalue with the smallest real part remains real throughout the entire parameter regime \cite{zhai2022nonequilibrium,zeng2025fidelity}. The system undergoes a transition from an extended phase to a localized phase, analogous to the Hermitian case. Importantly, the critical point coincides with the $\mathcal{PT}$ transition ~\cite{jiang2019interplay,zeng2025fidelity}. In the following, we demonstrate that under a power-law quasiperiodic potential, an additional mixed phase emerges. In this scenario, the $\mathcal{PT}$ transition occurs between these mixed states and the localized state, rather than between the extended and localized states.

\section{Methods}
\label{sec:methods}
In non-interacting disordered systems, two fundamental phases exist: the extended phase and the localized phase. In the extended phase, all states are delocalized, with wavefunctions spreading uniformly across the system. The spatial distribution follows \( |\psi(x_{i})|^2 \sim 1/L \), indicating a uniform probability of finding the particle throughout the lattice, where \( L \) is the system size. In contrast, in the localized phase, wavefunctions are centered around a position \( x_{0} \) and decay exponentially as \( |\psi(x_{i})| \sim \exp \left(-|x_{i} - x_{0}|/\xi \right) \), where \( \xi \) denotes the localization length.

The IPR is a widely used measure of wave-function localization in real space and serves as a key tool for studying localization-delocalization phase transitions. It is defined as
\begin{equation}
    \text{IPR}^{(m)}=\sum_{i=1}^{L}|\psi^{(m)}_i|^4,
\end{equation}
where $i$ labels the lattice sites, $m$ denotes the energy level, and $\psi_{i}^{(m)}$ is the $i$th component of the normalized single-particle eigenstate at level energy $m$. For extended states, the IPR is small and decreases toward zero as the system size $L$ increases, following the scaling relation $\text{IPR}\propto L^{-1}$. In contrast, for localized states, the IPR is large and saturates at a finite, nonzero value in the thermodynamic limit ($L \rightarrow\infty$), indicating that the particle is confined to a finite spatial region. Similarly, the normalized participation ratio (NPR) is defined as
\begin{equation}
    \text{NPR}^{(m)}=\left( L\sum_{i=1}^{L}|\psi^{(m)}_i|^4 \right)^{-1}.
\end{equation}
The NPR exhibits behavior opposite to that of the IPR: it takes a finite value for extended states and approaches zero for localized states as the system size increases. The fractal dimension is another valuable tool for distinguishing these states. It is extracted from the scaling of the generalized IPR \cite{evers2008anderson,banerjee2025emergence,roy2018multifractality,roy2022critical}, defined as
\begin{equation}
    \text{IPR}^{(m)}_q=\sum_{i=1}^{L}|\psi^{(m)}_i|^{2q}\propto L^{-\tau_q},
\end{equation}
where $\tau_q$ is the scaling exponent. The exponent $\tau_q$ is related to the fractal dimension $D_q$ via $\tau_q = D_q(q-1)$~\cite{banerjee2025emergence}. For extended states, $D_q = 1$, while for localized states, $D_q = 0$. Note that $\text{IPR}^{(m)}_q$ reduces to the standard IPR for $q=2$, which scales as $\text{IPR}_2 \propto L^{-D_2}$~\cite{banerjee2025emergence}.

In addition to the fractal dimension, another useful diagnostic for identifying the intermediate phase is given by~\cite{li2020mobility,roy2021reentrant,padhan2022emergence}
\begin{equation}
    \zeta = \log_{10} \left( \overline{\text{IPR}} \times \overline{\text{NPR}} \right),
\end{equation}
where $\overline{\text{IPR}} = \frac{1}{L}  \sum\limits_{m} \text{IPR}^{(m)}$ and $\overline{\text{NPR}} = \frac{1}{L} \sum\limits_{m} \text{NPR}^{(m)}$.
For the extended phase, the $\text{IPR}$ scales as \(1/L\) and the $\text{NPR}$ is of order unity, while the opposite behavior occurs in the localized phase \cite{li2020mobility,padhan2022emergence}. Consequently, \(\zeta \lesssim -\log_{10} L\)  in both the extended and localized regimes \cite{li2020mobility,padhan2022emergence}. In contrast, within the intermediate phase, both $\overline{\text{IPR}}$ and $\overline{\text{NPR}}$ are finite, resulting in a finite value of \(\zeta\) of order \(\mathcal{O}(1)\) \cite{li2020mobility,padhan2022emergence}.

In non-Hermitian systems, a key concept is $\mathcal{PT}$ symmetry, which gives rise to two distinct phases: the $\mathcal{PT}$-symmetric phase, in which all eigenvalues are purely real, and the $\mathcal{PT}$-broken phase, where eigenvalues appear as complex conjugate pairs. The transition between these phases occurs at an exceptional point (EP), where both eigenvalues and their corresponding eigenstates coalesce. To capture spectral changes across the $\mathcal{PT}$ transition, we introduce the density of states (DOS), defined as \cite{acharya2024localization}
\begin{equation}
    \text{DOS}=\frac{N(\text{Im}(E)\neq0)}{L},
\end{equation}
where $N$ denotes the number of eigenvalues with nonzero imaginary parts.

\begin{figure}[t]
\begin{center}
\includegraphics[width=8.7cm]{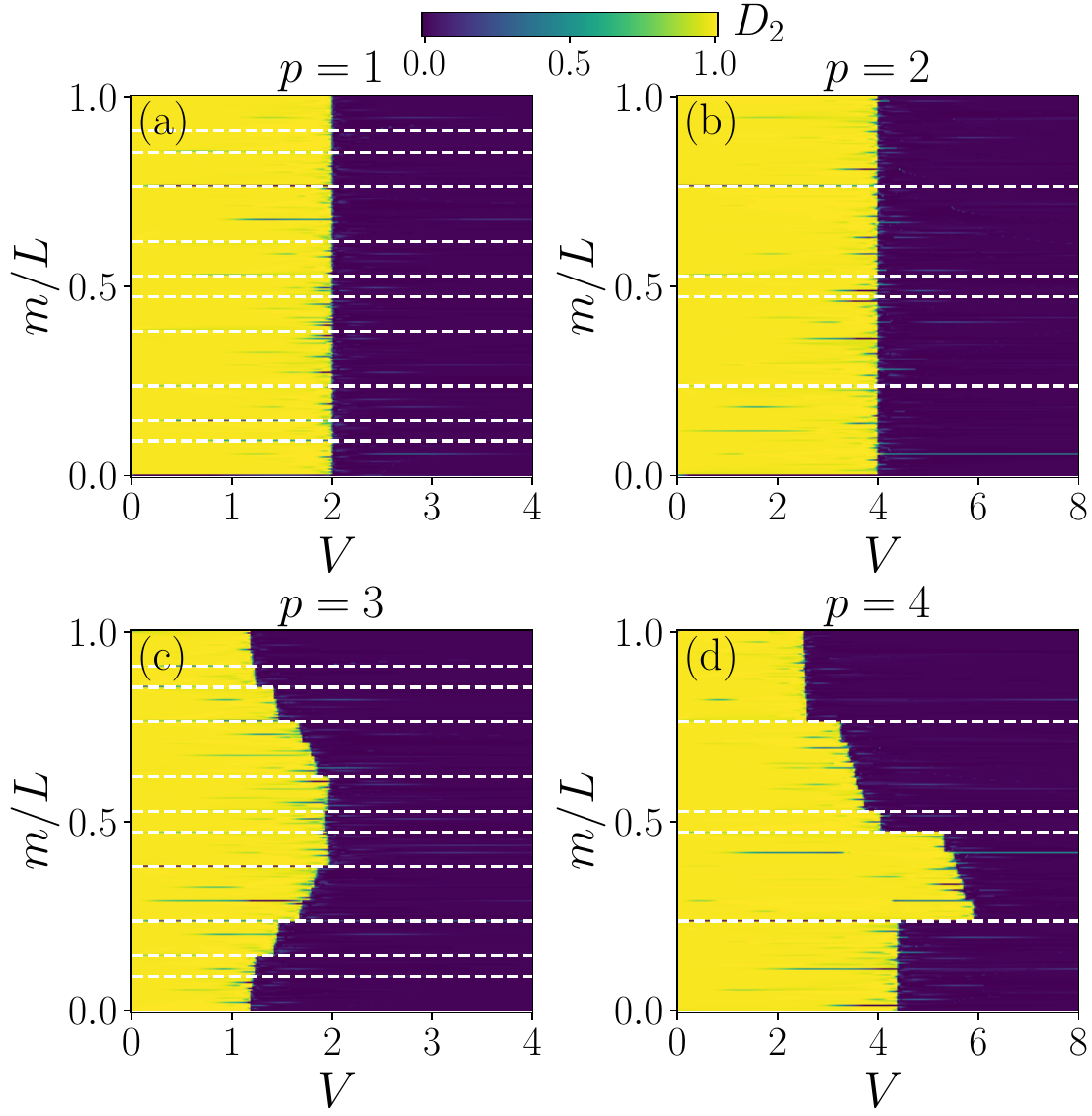} 
\caption{Fractal dimensions of the Hermitian AAH model in the thermodynamic limit. Panels (a)-(d) show the fractal dimensions $D_2$ of eigenstates for exponents $p = 1, 2, 3,$ and $4$, respectively. The white dashed lines in (a)-(d) indicate the locations of eigenstates exhibiting large IPR values as explained in the main text.}
\label{D2}
\end{center}
\end{figure}

\section{Results}
\label{sec:results}
The generalized AAH model, as given in Eq.~(\ref{eq:Ham}) , is solved using the Bogoliubov–de Gennes (BdG) formalism, \(\hat{H} = \mathbf{c}^\dagger \mathcal{H} \mathbf{c}\), where \(\mathbf{c} = (c_1, c_2, \dots, c_{L})^\text{T}\)  is the vector of fermionic annihilation operators on the lattice, and \(\mathcal{H}\) denotes the \(L \times L\) single-particle Hamiltonian matrix. Its explicit matrix form is given by
\[
\mathcal{H} =
\begin{pmatrix}
V(1) & -t e^{-g} & 0 & \cdots & -t e^{g} \\
-t e^{g} & V(2) & -t e^{-g} & \cdots & 0 \\
0 & -t e^{g} & \ddots & \ddots & 0 \\
\vdots & \vdots & \ddots & \ddots & -t e^{-g} \\
-t e^{-g} & 0 & 0 & -t e^{g} & V(L)
\end{pmatrix},
\]
where periodic boundary conditions are imposed. The Hamiltonian matrix is diagonalized to obtain the full spectrum and corresponding eigenstates, which are subsequently analyzed to explore the phase transitions in the generalized AAH model.

\subsection{Mobility Edge and Statistical Properties}
In the generalized AAH model, we first consider the Hermitian case ($g=0$) and choose the irrational number as the golden ratio, $\beta=(\sqrt{5}-1)/2$. The model reduces to the original AAH case when the modulation period is $p = 1$. In this scenario, it is well established that a transition from extended to localized phase occurs at the critical potential strength $V = 2$, without the emergence of mobility edges. This result is confirmed by numerical simulations, as shown in Fig.\ref{D2}(a). To evaluate the fractal dimension $D_2$ in the thermodynamic limit, we employ a finite-size scaling analysis. We calculate the average inverse participation ratio $\overline{\text{IPR}}$ over fixed intervals for various system sizes $L=1000, 2000, 3000, 4000$, and $5000$, and extract $D_2$ directly from the slope. When $V < 2$, all fractal dimensions approach unity, indicating an extended phase, whereas for $V > 2$, all fractal dimensions approach zero, signaling a localized phase. For $p=2$, the potential takes the form $V(i)=V\cos^2\theta_{i}=\frac{1}{2} V\left[ 1+ \cos(2\theta_{i}) \right]$, effectively reducing the potential amplitude to $V/2$, where $\theta_{i}=2\pi\beta i$. Since the self-dual transition point in the AAH model depends on the modulation amplitude, the phase transition correspondingly occurs at \( V = 4t \), as illustrated in Fig.\ref{D2}(b).

For $p=3$, the system exhibits markedly different behavior. Using the trigonometric identity $\cos^3\theta_{i} = \left[ 3\cos(\theta_{i}) + \cos(3\theta_{i}) \right]/4$, the quasiperiodic potential $V(i)$ can be interpreted as a superposition of two coupled frequency components. This coupling gives rise to a mixed phase, in which localized and extended states coexist in the energy spectrum, as illustrated in Fig.\ref{D2}(c). Similarly, for $p=4$, the potential $V(i) = V \left[ \cos(4\theta_{i})+4\cos(2\theta_{i})+3 \right]/8$ also comprises two coupled quasiperiodic components. The evolution from an extended phase to a mixed phase, and ultimately to a localized phase as $V$ increases, is confirmed by the fractal dimension shown in Fig.\ref{D2}(d).

\begin{figure}[t]
\centering
\includegraphics[width=9.3cm]{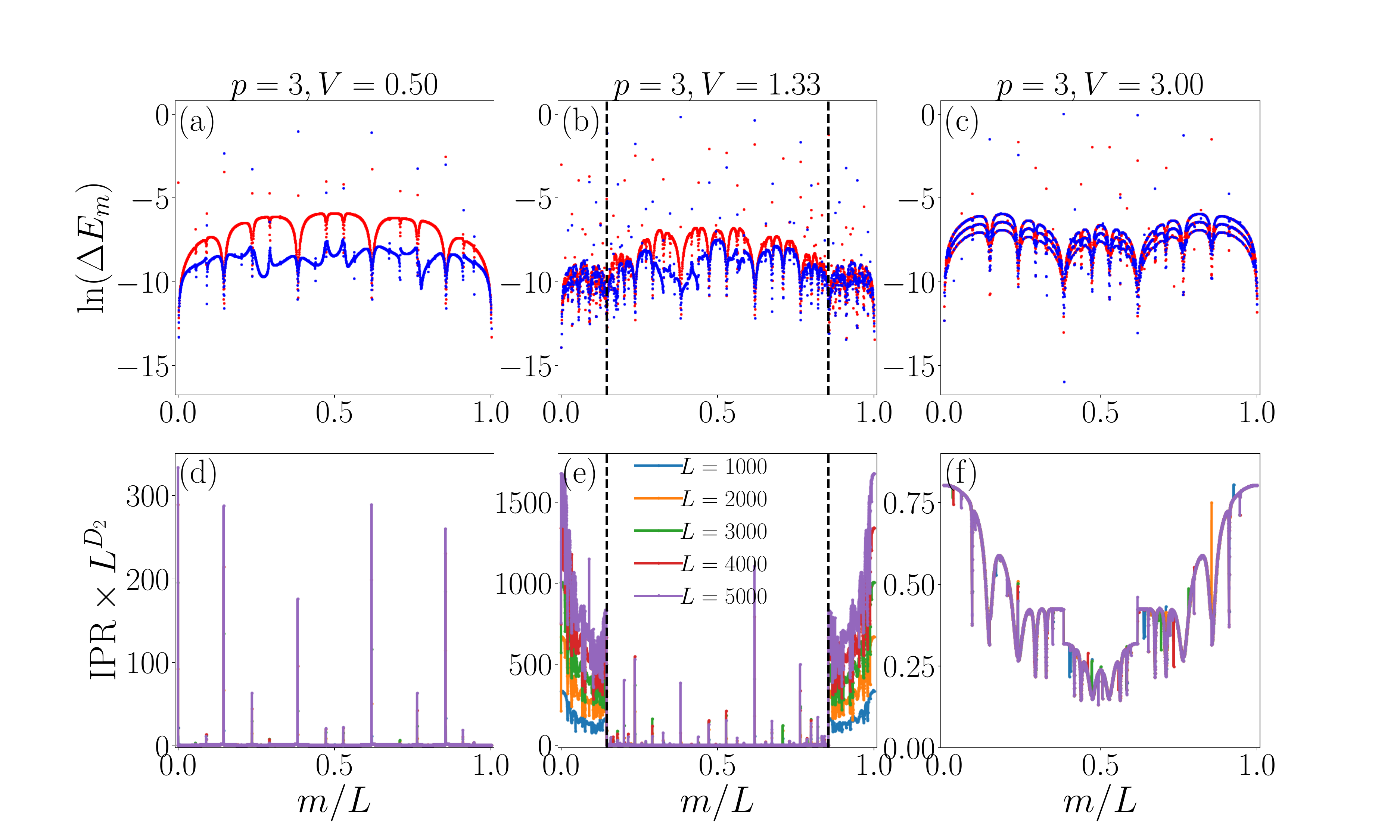} 
\caption{Energy differences in the $p = 3$ Hermitian AAH model with $L = 4000$. Panels (a)-(c) display the energy difference $\Delta E_{2m-1}=E_{2m}-E_{2m-1}$ (red dots) and $\Delta E_{2m}=E_{2m+1}-E_{2m}$ (blue dots) as functions of the eigenstate index $m/L$ for $V = 0.50$, $1.33$, and $3.00$ respectively. Panels (d)-(f) illustrate the data collapse of the IPR across different system sizes $L=1000, 2000, 3000, 4000,$ and $5000$. The specific phases and their corresponding fractal dimensions used for scaling are: (d) extended phase ($V=0.50$, $D_2=1$), (e) intermediate phase ($V=1.33$, $D_2=1$), and (f) localized phase ($V=3.00$, $D_2=0$). In panel (b) and (e), the two black dashed lines mark the positions of the mobility edges at $m/L = 0.14575$ and $0.8545$.}
\label{E}
\end{figure}

To distinguish different phases and gain deeper insight into the eigenstate behavior in the mixed phase, we compute the energy level spacings for $p=3$, as shown in Fig.\ref{E}. The energy differences are defined as~\cite{deng2019one,sarkar2021mobility,zhang2022localization} $\Delta E_{2m-1} = E_{2m} - E_{2m-1}$ and $\Delta E_{2m} = E_{2m+1} - E_{2m}$ with the eigenvalues $E_{2m}$ and $ E_{2m-1}$ arranged in ascending order. In the extended phase ($V=0.5$), the energy spectrum is nearly doubly degenerate, resulting in $\Delta E_{2m}\approx 0$ corresponding to small negative values of $\ln(\Delta E_{2m})$, and producing a pronounced gap between $\ln(\Delta E_{2m-1})$ and $\ln(\Delta E_{2m})$, as shown in Fig.\ref{E}(a). In the localized phase ($V=3$), $\ln(\Delta E_{2m})$ and $\ln(\Delta E_{2m-1})$ behave similarly, closing this gap, as illustrated in Fig.\ref{E}(c). In contrast, in the mixed phase ($V=1.33$), the eigenstates are localized in the regimes $m/L<0.14575$ and $m/L>0.8545$, while remaining extended in the intermediate regime [see Fig.\ref{E}(b)]. The phase characteristics are further confirmed through the data collapse of the IPR across various system sizes $L=1000, 2000, 3000, 4000$, and $5000$ as shown in Figs. \ref{E}(d)-\ref{E}(f). Specifically, for the extended phase $V=0.5$, the scaled quantity $\text{IPR} \times L^{D_2}$ exhibits an excellent data collapse with fractal dimension $D_2=1$. In the mixed phase $V=1.33$, the same scaling $D_2=1$ leads to a clear collapse only within the extended intermediate regime, which separates the localized states located at the two spectral edges. In contrast, for the localized phase at $V=3$, all eigenstates are strongly localized, resulting in a data collapse with $D_2=0$.

\begin{figure}[t]
\centering
\includegraphics[width=8.6cm]{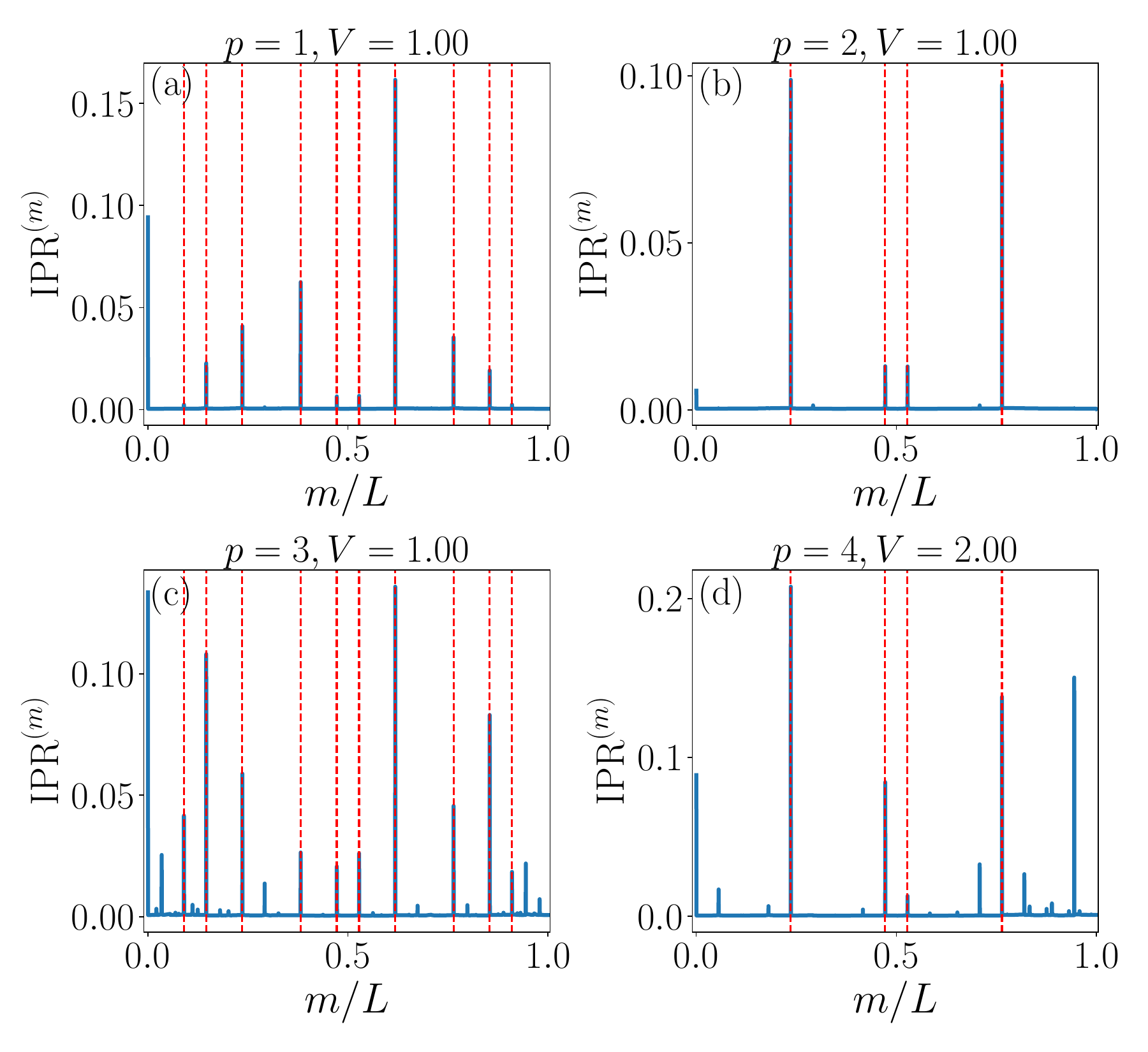}
\caption{The $\text{IPR}^{(m)}$ of Hermitian AAH model as a function of the eigenstate index $m/L$ for different values of $p$ and $V$ at $L = 4000$. (a) $p = 1$, $V = 1$; (b) $p = 2$, $V = 1$; (c) $p = 3$, $V = 1$; (d) $p = 4$, $V = 2$. The distinct vertical lines indicate the positions of eigenstates with high IPR values, while the red dashed lines indicate the corresponding theoretical predictions for large IPR values.}
\label{IPRfig}
\end{figure}

Recent studies have demonstrated that the $p=1$ AAH model can host high-IPR states, which occur at the positions of spectral gaps \cite{roy2019study,roy2021fraction}. The corresponding irrational numbers are shown to satisfy a universal relation governed by a nontrivial rule that determines the locations of all high-IPR states \cite{roy2019study,roy2021fraction}. It is known that for a rational $\beta = q_r/q_s$, the positions of the energy gap satisfy the Diophantine relation \cite{thouless1982quantized,kohmoto1987critical,kohmoto1992diophantine}, now known as Thouless-Kohmoto-Nightingale-den Nijs (TKNN) equation, given by $x=n_s q_s + n_r q_r$, where $n_s$ and $n_r$ are integers. For an irrational $\beta$ (such as the golden, silver, or bronze ratio), the spectrum forms a Cantor set \cite{kohmoto1987critical,kohmoto1992diophantine,Sarkar2023Tuning}, however the gap positions remain located at $x/q_s = n_s + n_r \beta$ \cite{lang2012edge}. We argue that high-IPR states can emerge at gap positions in the generalized AAH model for all values of $p$ and irrational $\beta$, as illustrated in Fig.\ref{IPRfig} and Appendix~\ref{AppNA} and \ref{AppA}, and that their positions (modulo one) obey the same simple universal relation
\begin{equation}
\label{na-na}
    x_{n}=n\beta-\lfloor n\beta \rfloor,
\end{equation}
Here, $x_n$ denotes the positions of the high-IPR states, $n$ is a non-negative integer, and $\lfloor \cdot \rfloor$ denotes the floor function. The resulting positions $x_n$ exhibit a mirror symmetry, satisfying $x_n + x_n^{\prime} = 1$ with respect to the midpoint $x_n=1/2$. The value of $n$ is determined by the form of the potential: for a $\cos\theta_{i}$ potential, $n$ takes positive integer values $\{1, 2, 3, ...\}$; for a $\cos2\theta_{i}$ potential, $n$ takes even values $\{2, 4, 6, ...\}$; and for a $\cos3\theta_{i}$ potential, $n$ follows the sequence of multiples of three, $\{3, 6, 9, ...\}$. Accordingly, for a multi-frequency coupled potential such as $\cos^3\theta_{i}$, which can be expanded into a linear combination of $\cos\theta_{i}$ and $\cos3\theta_{i}$, the resulting $n$ sequence spans all natural numbers $\{1, 2, 3, ...\}$.

It is worth noting that the energy spectrum of the $p=3$ AAH model exhibits a stair-like hierarchical structure. The positions of the energy levels for high-IPR states accurately follow the relation in Eq.(\ref{na-na}), thereby predicting the mobility edges. We numerically verify this relation for $p=3,V=1$ [see Fig.\ref{IPRfig}(c)] and $p=4,V=2$ [see Fig.\ref{IPRfig}(d)] by calculating the IPR of each eigenstate. The locations of the high-IPR states in Figs.\ref{IPRfig}(c) and \ref{IPRfig}(d), coincide with the white dashed lines in Figs.\ref{D2}(c) and \ref{D2}(d). A comparison between the numerical and theoretical positions of the high-IPR states for $p=3$ and $p=4$ is presented in Table~\ref{Goldtable} based on the parameter $n$ and the golden irrational number. The deviations are on the order of $10^{-4}$. Additionally, we have verified that this relation also holds for the silver and bronze ratio irrational numbers [see Appendix~\ref{AppA}]. This finding demonstrates that the positions of high-IPR states can be accurately predicted, providing a unified framework for understanding the spatial distribution of states and the corresponding energy levels in quasiperiodic systems. We note that the prediction of the positions of high-IPR states in Eq. (\ref{na-na}) is meaningful mainly for low-order $n$ associated with well-resolved spectral gaps. For large $n$, the sequence becomes dense in $[0,1]$ and the corresponding gaps become exceedingly small, making precise numerical identification impractical.

\begin{table}[t]
\centering
\caption{Numerical and theoretical results for the positions of high IPR in the generalized AAH model for $p=3$, $V=1$ and $p=4$, $V=2$, respectively. }
\label{Goldtable}
\begin{tabular}{|c|c|c|c|c|}
\hline
\multicolumn{5}{|c|}{$p = 3,V=1$} \\
\hline
\multirow{2}{*}{$n$} & \multicolumn{2}{c|}{$x_n$} & \multicolumn{2}{c|}{$x_n'$} \\
\cline{2-5}
 & Numerical Value & Theory & Numerical Value & Theory \\
\hline
1 & 0.382 & 0.381966 & 0.61825 & 0.618034 \\
\hline
2 &0.23625 & 0.236068 & 0.764 &0.763932 \\
\hline
3 & 0.14575 & 0.145898 & 0.8545 & 0.854102 \\
\hline
4 & 0.47225 & 0.472136 & 0.528 & 0.527864 \\
\hline
5 & 0.09025 & 0.090170 & 0.91000 & 0.909830 \\
\hline
\multicolumn{5}{|c|}{$p = 4,V=2$} \\
\hline
\multirow{2}{*}{$n$} & \multicolumn{2}{c|}{$x_n$} & \multicolumn{2}{c|}{$x_n'$} \\
\cline{2-5}
 & Numerical Value & Theory & Numerical Value & Theory \\
\hline
2 & 0.23625 & 0.236068 & 0.76425 & 0.763932 \\
\hline
4 & 0.47225 & 0.472136 & 0.528 & 0.527864 \\
\hline
\end{tabular}
\end{table}

\subsection{$\mathcal{PT}$-Symmetry Breaking and Phase Diagram}
\label{sec:NH}
\begin{figure}[t]
\includegraphics[width=8.7cm]{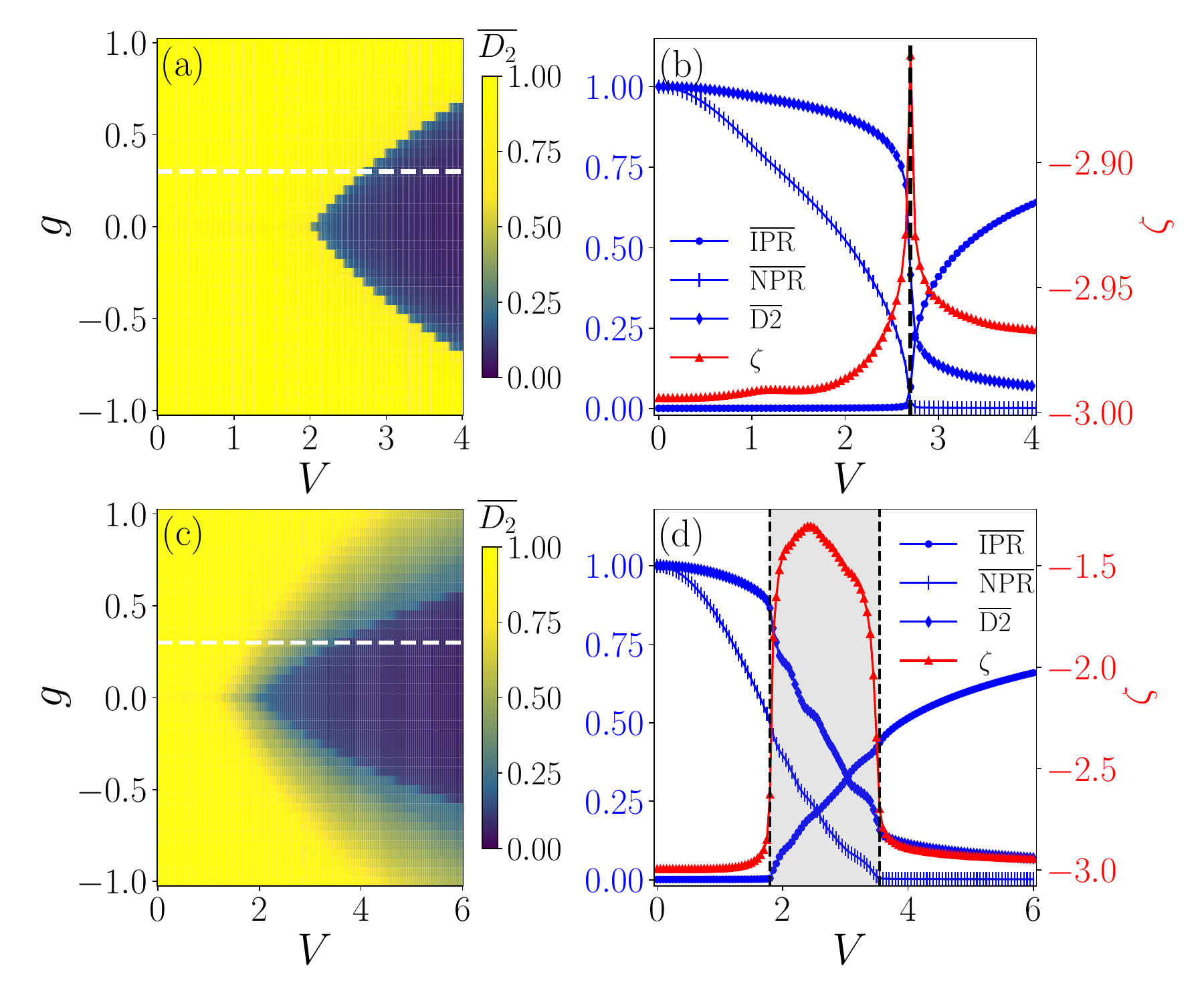} \centering
\caption{Phase diagram of the non-Hermitian AAH model.
(a) Phase diagram obtained from the fractal dimension $\overline{D}_2$ for $L = 987$ and $p = 1$; the white dashed line denotes $g = 0.3$.
(b) $\overline{\text{IPR}}$ , $\overline{\text{NPR}}$, $\overline{D}_2$, and $\zeta$ as functions of $V$ at $g = 0.3, p=1$; the left vertical axis corresponds to $\overline{\text{IPR}}$, $\overline{\text{NPR}}$, and $\overline{D}_2$ while the right vertical axis corresponds to $\zeta$, The black dashed line marks the phase transition from extended to localized states at $V_c = 2.7$.
(c) Phase diagram based on $\overline{D}_2$ for $L = 987$, $p = 3$; the white dashed line indicates $g = 0.3$.
(d) $\overline{\text{IPR}}$ , $\overline{\text{NPR}}$, $\overline{D}_2$, and $\zeta$ as functions of $V$ at $g = 0.3, p=3$; the gray-shaded region $1.8 < V < 3.55$ denotes the mixed phase where extended and localized states coexist.}
\label{MIPRfig}
\end{figure}

We next investigate a generalized non-Hermitian AAH model subject to a power-law quasiperiodic potential. When $p=1$, the Hamiltonian in Eq.(\ref{eq:Ham}) reduces to the standard nonreciprocal AAH model~\cite{jiang2019interplay,zeng2025fidelity}. Previous studies~\cite{jiang2019interplay,zeng2025fidelity} have shown that the system undergoes a direct transition from the extended phase to the localized phase as the potential strength increases, with the $\mathcal{PT}$ transition coinciding exactly with the extended-localized phase transition point. Consequently, the energy spectrum is complex in the extended phase and purely real in the localized phase.
We present the full phase diagram based on the fractal dimension $\overline{D}_2$ from the right eigenvectors, as shown in Fig.\ref{MIPRfig}(a) for $p=1$ with $L=987$. Figure \ref{MIPRfig}(b) illustrates the behavior of $\overline{\text{IPR}}$, $\overline{\text{NPR}}$, $\overline{D_2}$ and $\zeta$ as functions of the potential strength $V$, supporting the phase diagram. A vanishing $\overline{\text{IPR}}$ characterizes the extended phase, while a vanishing $\overline{\text{NPR}}$ identifies the localized phase. In contrast, a large value of $\zeta$ indicates the presence of a mixed phase. The critical point at $V=2.7$, denoted by the black dashed line in Fig.\ref{MIPRfig}(b), is consistent with the theoretically predicted phase boundary $2e^g$ reported in previous studies \cite{jiang2019interplay,zeng2025fidelity}. For $p=2$, the system behaves similarly to the $p=1$ case. In contrast, for $p \geq 3$, the system exhibits distinctly different behavior. The phase diagram, shown in Fig.\ref{MIPRfig}(c), is obtained from the fractal dimension $\overline{D}_2$ for $p=3$ with $L=987$, where the light-green region clearly indicates the presence of an intermediate mixed phase. To characterize the extended, mixed, and localized phases, we calculate $\overline{\text{IPR}}$, $\overline{\text{NPR}}$, $\overline{D_2}$ and $\zeta$ for $L=987$ at $g=0.3$, as shown in Fig.\ref{MIPRfig}(d). It is observed that in the range $1.8<V<3.55$, both $\overline{\text{IPR}}$ and $\overline{\text{NPR}}$ remain finite, revealing the coexistence of extended and localized states. Meanwhile, $\zeta$ exhibits pronounced nonzero values, confirming the emergence of an intermediate mixed phase. We note that the phase diagram does not depend on whether the system size is a Fibonacci number.

\begin{figure}[t]
\includegraphics[width=8.3cm]{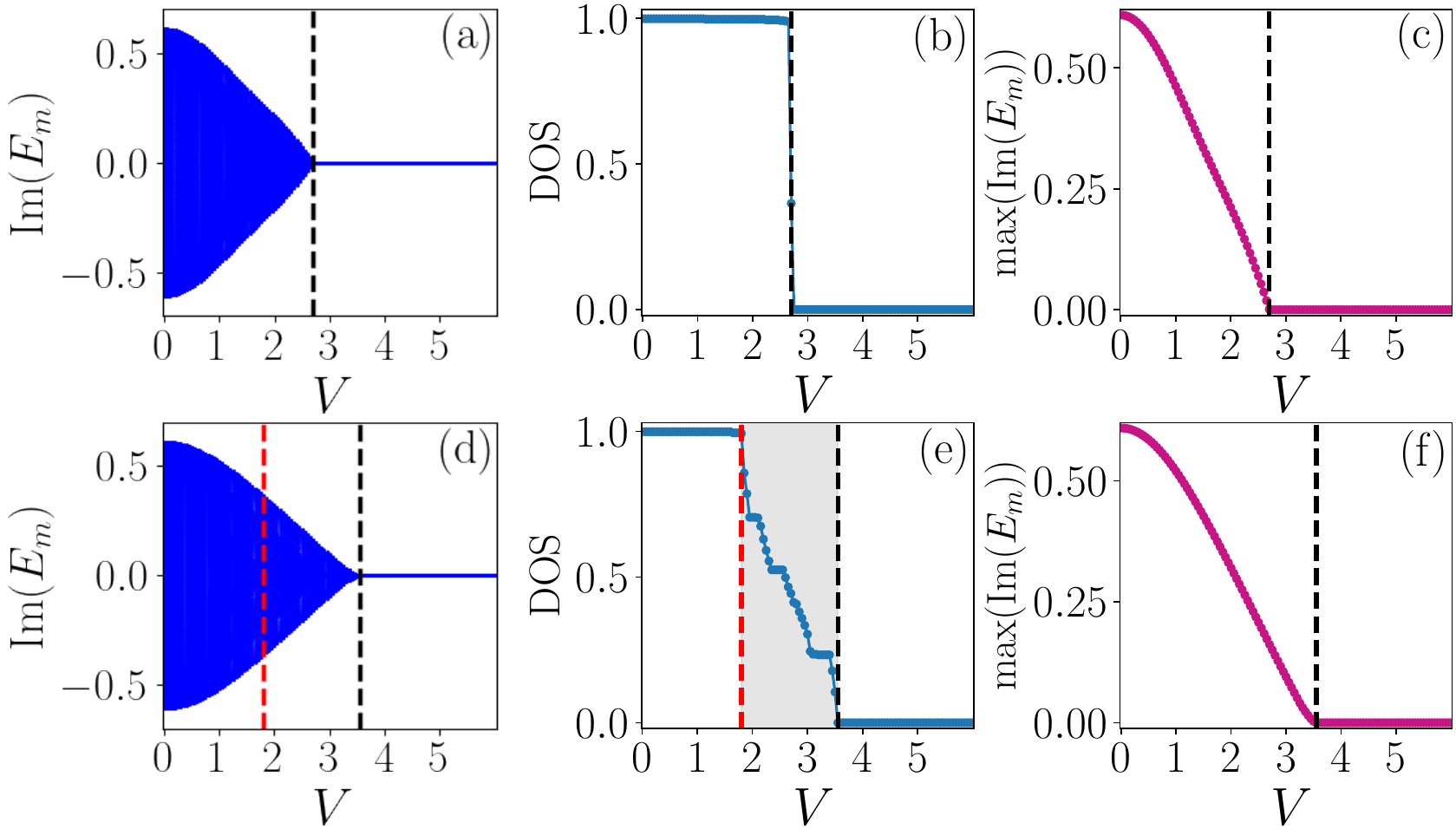} 
\centering
\caption{Spectral properties of the non-Hermitian AAH model at $L = 987$ and $g = 0.3$. Panels (a)-(c) correspond to $p = 1$; (a) imaginary parts of the eigenvalues $\operatorname{Im}(E_m)$ as a function of $V$; (b) density of states (DOS) versus $V$; (c) maximum imaginary part of the eigenvalues $\max(\operatorname{Im}(\text{E}_\text{m}))$ with respect to $V$. In panels (a)-(c), the black dashed line indicates the \(\mathcal{PT}\) transition at $V = 2.7$. Panels (d)-(f) show the same quantities as in (a)(b)(c) at $p=3$, where the red and black dashed lines mark the extended-to-mixed state transition and the \(\mathcal{PT}\) transition at $V = 1.8$ and $V = 3.55$, respectively.}
\label{ImE}
\end{figure}

Nonreciprocal hopping gives rise to complex energies. To investigate the spectral properties and their connection to the phase transitions shown in Fig.\ref{MIPRfig}, we present the full energy spectra in Fig.\ref{ImE}. For $p=1$, the spectrum undergoes a direct \(\mathcal{PT}\) transition at \( V = 2.7 \), as shown in Figs.\ref{ImE}(a)-(c), which coincides exactly with the extended-to-localized phase transition.
For \( p = 3 \), the system still exhibits a direct \(\mathcal{PT}\) transition at \( V=3.55 \), as shown in Fig.\ref{ImE}(d). However, as $V$ increases, the system evolves not from an extended phase with predominantly imaginary eigenvalues to a localized phase, but into a mixed phase comprising both real and imaginary eigenvalues [see Fig.\ref{ImE}(e)]. With further increase of $V$, the system ultimately transitions to a localized phase with entirely real eigenvalues, corresponding to a \(\mathcal{PT}\) transition between the mixed and localized phases [see Fig.\ref{ImE}(f)].
The critical points  \( V = 1.8 \) and \( V = 3.55 \), indicated by the red and black dashed lines in Fig.\ref{ImE}(d) and (e) are consistent with the boundaries of the gray region shown in Fig.\ref{MIPRfig}(d). Similar to the \( p = 3 \), the $p=4$ non-Hermitian AAH model exhibits a \(\mathcal{PT}\) transition that aligns with the mixed-to-localized transition rather than the extended-to-localized transition [see Appendix~\ref{AppB}]. This finding highlights the interplay between non-Hermiticity and quasiperiodicity, providing a unified framework for exploring novel phases and phase transitions in non-Hermitian quasiperiodic systems.

\section{Wave-packet Spreading}
\label{sec:wave}

\begin{figure*}[htbp]
\includegraphics[width=15cm]{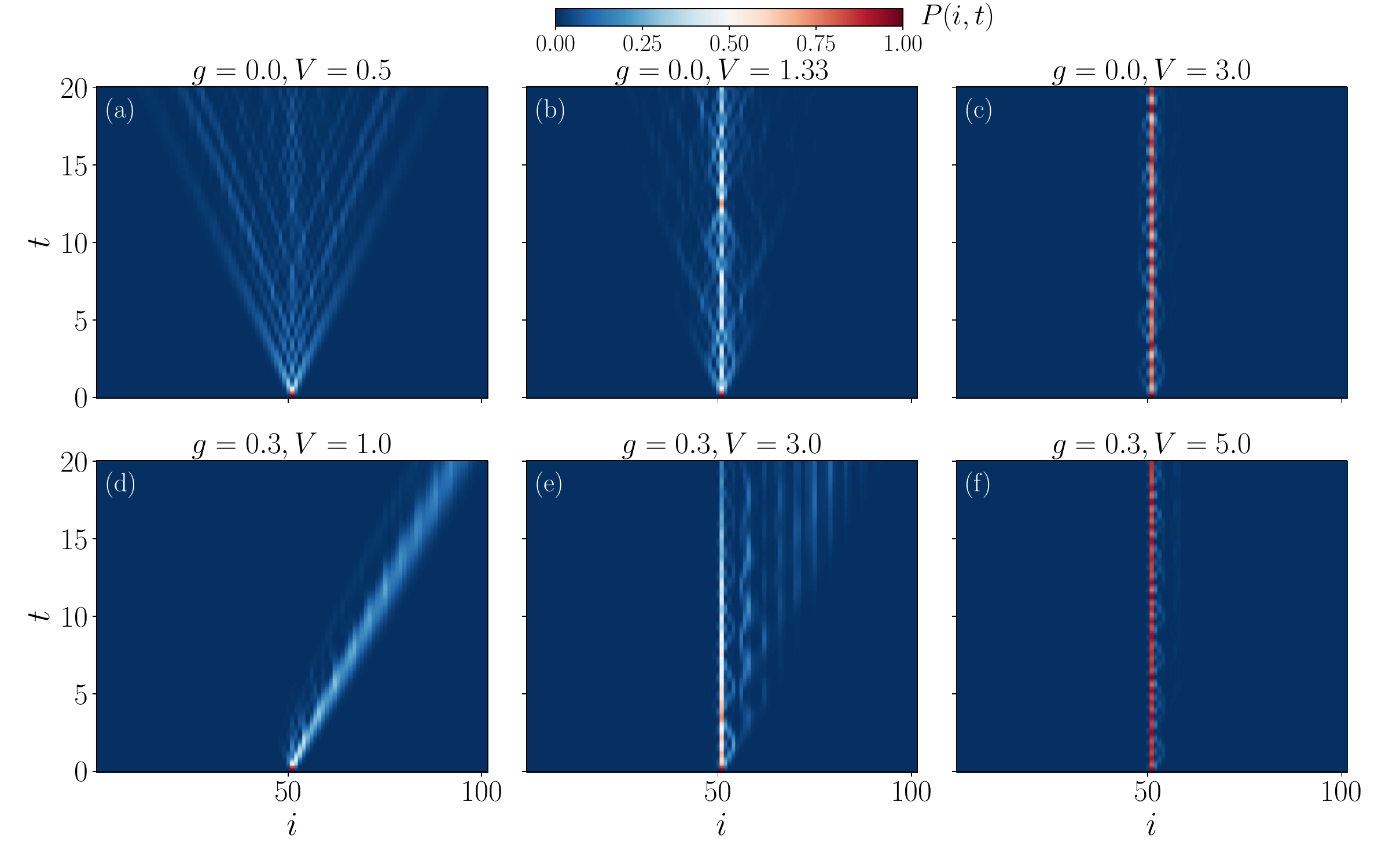}
\centering
\caption{Spatiotemporal evolution of the probability distribution $P(i,t)$ for a single-particle wave packet with $p=3$ at $\beta=(\sqrt{5}-1)/2$. The system size is $L=101$, and the initial state is $|\psi(0)\rangle = |i_{51}\rangle$, localized at the central site. Panels (a)-(c) correspond to the Hermitian case ($g=0$). In (a), within the extended phase ($V=0.5$), the wave packet exhibits symmetric, ballistic spreading. In (b), within the mixed regime ($V=1.33$), anomalous spreading arises due to the presence of a mobility edge. In (c), within the localized phase ($V=3.0$), the probability distribution remains confined near the initial site. Panels (d)-(f) correspond to the non-Hermitian case ($g=0.3$). In (d), within the extended phase ($V=1.0$), the wave packet displays unidirectional transport toward the right boundary, a hallmark of the non-Hermitian skin effect (NHSE). In (e), within the intermediate regime ($V=3.0$), the system exhibits asymmetric mixed spreading. In (f), within the localized phase ($V=5.0$), the strong quasiperiodic potential dominates over the non-Hermitian driving, effectively pinning the particle near its initial position.}
\label{wavepacket.fig}
\end{figure*}

In this section, we investigate the transport properties of the system by simulating the dynamical evolution of a single-particle wave packet. We consider a system of size $L=101$ with $p=3$, and choose the initial state to be fully localized at the central lattice site $i_0 = 51$, i.e., $|\psi(0)\rangle = |i_{51}\rangle$. The time evolution of the system is governed by the Hamiltonian $H$, yielding the state $|\psi(t)\rangle$. The spatiotemporal evolution is characterized by the probability distribution at each site, $P(i,t) = |\langle i |\psi(t) \rangle|^2$.

As shown in Figs.~\ref{wavepacket.fig}(a)-(c), the probability distribution exhibits pronounced symmetry in the Hermitian case ($g=0$). In the extended phase ($V=0.5$), the wave packet undergoes rapid, symmetric spreading, with the probability density expanding from the center toward both boundaries until the wavefront reaches the edges of the lattice, reflecting the high transport efficiency of this phase [cf. Fig.~\ref{wavepacket.fig}(a)]. In the localized phase ($V=3.0$), due to Anderson localization induced by the quasiperiodic potential, the wave packet shows negligible spreading, and the probability distribution remains strongly confined near the initial site $i=51$ throughout the evolution [cf. Fig.~\ref{wavepacket.fig}(c)]. In the mixed regime ($V=1.33$), the system exhibits anomalous diffusion, with both the spatial extent and intensity of spreading significantly reduced compared to the extended phase. This behavior arises from the presence of a mobility edge in the energy spectrum, which suppresses global transport [cf. Fig.~\ref{wavepacket.fig}(b)].

As shown in Figs.~\ref{wavepacket.fig}(d)-(f), the presence of non-Hermiticity ($g=0.3$) induces asymmetric hopping amplitudes, which break the symmetry of the dynamics. In the extended phase ($V=1.0$), unidirectional propagation is observed. Due to the stronger rightward hopping, the wave packet shifts predominantly to the right and rapidly accumulates at the boundary, providing clear evidence of the non-Hermitian skin effect (NHSE) [cf. Fig.~\ref{wavepacket.fig}(d)]. In the localized phase ($V=5.0$), despite the asymmetric hopping induced by non-Hermiticity, the quasiperiodic potential dominates, and the particle remains strongly confined near the initial site [cf. Fig.~\ref{wavepacket.fig}(f)]. In the intermediate regime ($V=3.0$), the system exhibits asymmetric mixed spreading, where the wave packet displays sub-ballistic spreading accompanied by a rightward drift induced by non-Hermiticity.

\section {Conclusion}
\label{sec:Con}
In summary, we have investigated a generalized non-Hermitian AAH model with a power-law quasiperiodic potential, uncovering rich physical phenomena that go beyond the scope of the conventional AAH model. In the Hermitian case, for power-law exponents \( p \geq 3 \), the system hosts an intermediate mixed phase separating extended and localized phases, accompanied by the emergence of mobility edges. Notably, for system sizes that are not Fibonacci numbers, the IPR exhibits a series of pronounced peaks at specific energy levels. The positions of these high-IPR states are accurately captured by the analytical relation Eq. \eqref{na-na}, where the index $n$ is determined by the exponent $p$. In the non-Hermitian regime, the energy spectrum becomes complex. For \( p = 1, 2 \), the system undergoes a direct transition from the extended to localized phases, with the \(\mathcal{PT}\) transition coinciding with the extended-localized phase boundary. However, for \( p \geq 3 \), an intermediate mixed phase emerges, characterized by the coexistence of extended and localized states. In this case, the \(\mathcal{PT}\) transition aligns with the boundary between the mixed and localized phases, rather than the extended-localized transition. Our results uncover a rich interplay between non-Hermiticity and quasiperiodic modulation, paving the way for future investigations into topological transport and quantum control in non-Hermitian systems.

\begin{acknowledgments}
G.S. is appreciative of support from the NSFC under the Grants No. 11704186, "the Fundamental Research Funds for the Central Universities, No. NS2023055". W.-L. You acknowledges support from the NSFC under Grant No. 12174194. Y.-N.W. is supported by the Postgraduate Research {\&} Practice Innovation Program of Jiangsu Province under No.KYCX24\_0526. Z.X. is supported by the NSFC (Grant No. 12375016 and No. 12461160324), and Beijing National Laboratory for Condensed Matter Physics (No. 2023BNLCMPKF001). 
\end{acknowledgments}

The data that support the findings of this article are openly available \cite{opendata}.

\bibliography{ref}

\clearpage
\begin{widetext}
\appendix
\setcounter{figure}{0}
\renewcommand{\thefigure}{A\arabic{figure}}
\section{Spectral gaps and High-IPR states}
\label{AppNA}

In the extended phase, the high-IPR states originate from the unique mathematical structure of the single-particle energy spectrum in the generalized AAH model. For an irrational $\beta$, the energy spectrum forms a characteristic Cantor set, a fractal structure consisting of infinitely many gaps rather than continuous intervals. The positions of these energy gaps satisfy a Diophantine relation. Within this framework, numerous band gaps host high-IPR states, whose positions follow the expression given in Eq. (\ref{na-na}) in terms of the normalized level index.

As illustrated in the top panels of Figs. \ref{E_Delta.fig}(a)-(d), we show the evolution of the energy as a function of the normalized eigenstate index $m/L$ for $L = 4000$ and various power-law exponents ($p = 1, 2, 3, 4$). The step-like discontinuities clearly reflect the fragmented nature of the spectrum. Correspondingly, the bottom panels (e)-(h) present the energy differences as a function of the normalized eigenstate index, where the level spacings exhibit pronounced peaks at specific positions. This distinctive feature, which highlights the underlying spectral structure, is particularly evident for non-Fibonacci system sizes and persists as the system size increases. Since these states are located at spectral gap positions, the associated wave functions cannot extend into neighboring subbands and are thus effectively confined as high-IPR states. As a result, they display localization characteristics that differ markedly from the surrounding extended states.

\begin{figure}[hbp]
\includegraphics[width=16cm]{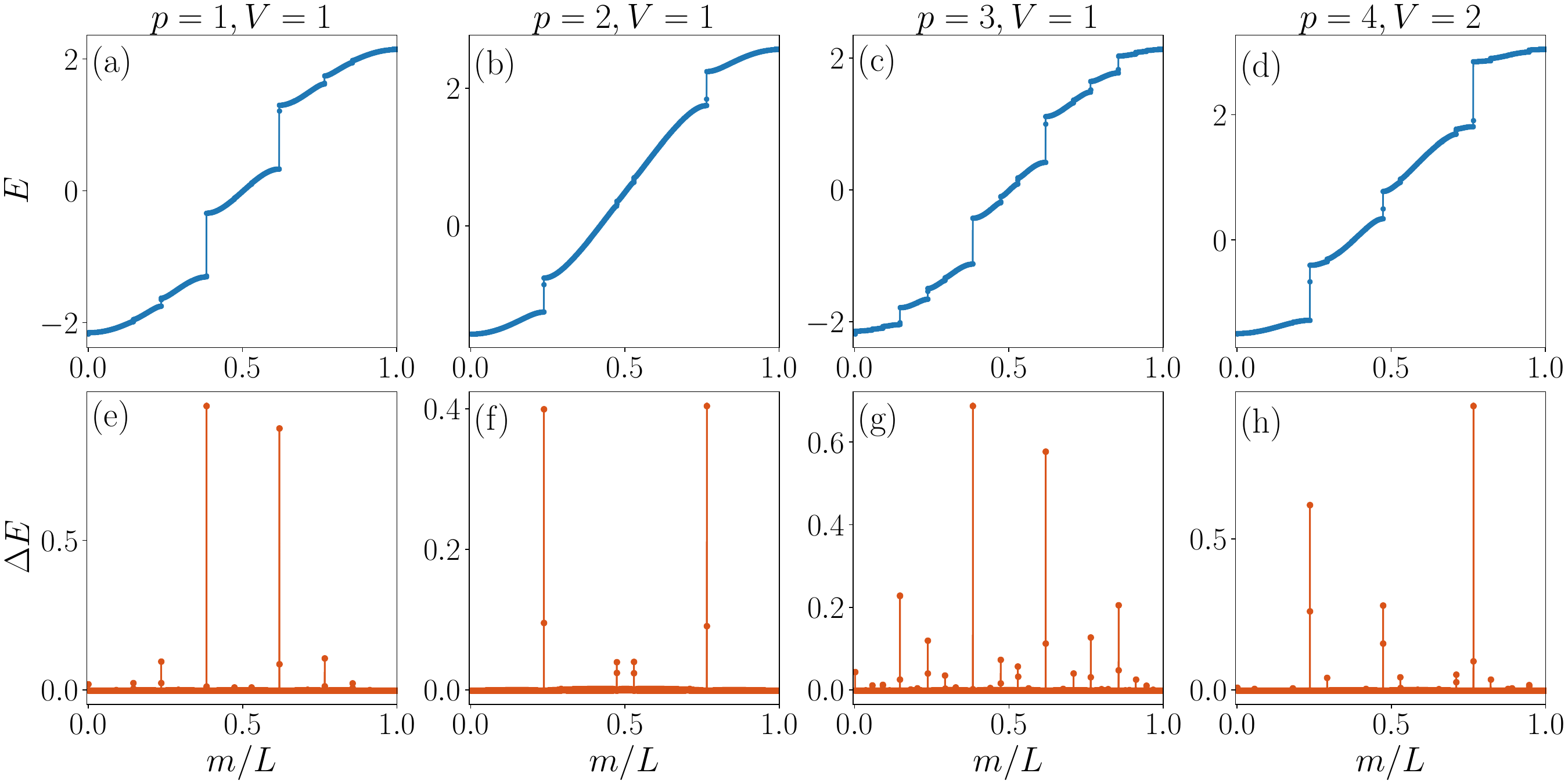}
\centering
\caption{Spectral structure and energy differences of the generalized AAH model at $g = 0$ and $\beta=(\sqrt{5}-1)/2$ for various power-law exponents $p$. Numerical results are shown for four representative parameter sets ($p = 1$, $V = 1$; $p = 2$, $V = 1$; $p = 3$, $V = 1$; and $p = 4$, $V = 2$) with system size $L = 4000$. The top panels (a)-(d) display the single-particle energy eigenvalues as a function of the normalized eigenstate index $m/L$, where the spectra exhibit pronounced step-like discontinuities. The bottom panels (e)-(h) show the variations of the adjacent energy differences ($E_{m+1} - E_m$) as a function of $m/L$, where significant peaks in the level spacing appear at positions corresponding to the spectral gaps in the top panels. In all subplots, the normalized eigenstate index $m/L$ is ordered in ascending energy.}
\label{E_Delta.fig}
\end{figure}

\setcounter{figure}{0}
\renewcommand{\thefigure}{B\arabic{figure}}
\section{Scaling analysis of the fractal dimension}
\label{AppNB}

\begin{figure}[htp]
\includegraphics[width=18cm]{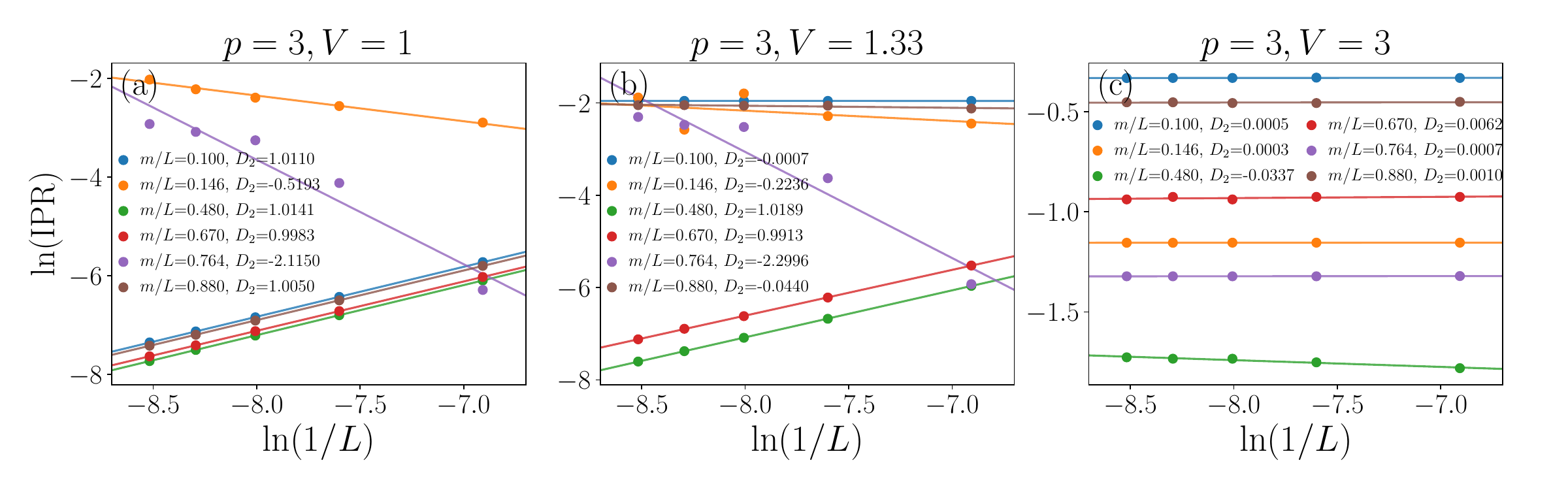}
\centering
\caption{Scaling of the IPR with system size $L$ for $p=3$ in the Hermitian case ($g=0,\beta=(\sqrt{5}-1)/2$). The numerical data are calculated for system sizes $L = 1000, 2000, 3000, 4000,$ and $5000$. The fractal dimension $D_2$ is extracted from the linear fit of $\ln(\text{IPR})$ versus $\ln(1/L)$. (a) Extended phase ($V=1$): the majority of eigenstates exhibit $D_2 \approx 1$ (e.g., $D_2 = 1.0110$ or $0.9983$), while special states at high-IPR positions (e.g., $m/L = 0.146$ and $0.764$) show significant deviations from power-law scaling, rendering them unsuitable for fitting. (b) Mixed phase ($V=1.33$): the coexistence of extended states ($D_2 \approx 1$) and localized states ($D_2 \approx 0$) is clearly observed due to the presence of a mobility edge. (c) Localized phase ($V=3$): the system enters a fully localized regime, with the fitted fractal dimension $D_2$ of all sampled states approaching zero (e.g., $D_2 = 0.0010$ at $m/L = 0.880$).}
\label{slope.fig}
\end{figure}

In this section, we quantitatively extract the fractal dimension $D_2$ in the thermodynamic limit through finite-size scaling analysis. 
According to the relation $\text{IPR}_2 \propto L^{-D_2}$, we take the logarithm of both sides and perform a linear fit between $\ln(\text{IPR}_2)$ and $\ln(1/L)$, where the fitted slope is identified as the fractal dimension $D_2$ corresponding to that energy level. For the Hermitian AAH model ($g=0$) with a power-law exponent $p=3$, we calculate the IPR for system sizes $L=1000$, $2000$, $3000$, $4000$, $5000$. Since the energy levels differ for different system sizes, we fix the normalized eigenstate index $m/L$ and perform linear fitting across different regimes. From this analysis, one can clearly distinguish the different behaviors in the extended and localized phases as shown in Fig.~\ref{slope.fig}.

In the extended phase ($V=1$), the fractal dimensions of the eigenstates, except for a small number of high-IPR states, are close to unity ($D_2 \approx 1$) [c. f. Fig.~\ref{slope.fig}(a)], indicating clear delocalization. For example, at $m/L=0.880$, the fitted value $D_2 = 1.0050$, shows that the corresponding wave function is well extended over the entire system. However, for a few high-IPR states, no clear linear scaling relation is observed, making it impossible to extract a physically meaningful value of $D_2$. A representative example occurs at $m/L=0.764$, where the fitted $D_2$ takes an unphysical negative value of $-2.1150$. 
We note that, for these isolated gap-edge high-IPR states, the finite-size scaling of individual eigenstates is not smooth because the corresponding spectral gaps and nearby levels shift with system size. Therefore, the fitted slope should not be interpreted as a physical fractal dimension. These isolated states are excluded from the phase identification based on the averaged scaling behavior.
As the system enters the mixed phase ($V=1.33$), the energy spectrum exhibits a hierarchical structure, in which extended and localized states coexist. Accordingly, both $D_2=1$ and $D_2=0$ are observed, corresponding to extended and localized states, respectively [c. f. Fig.~\ref{slope.fig}(b)]. For instance, in the localized region (i.e. $m/L=0.100$), the fractal dimension is approximately $-0.0007$, while in the extended region (i.e. $m/L=0.480$), $D_2=1.0189$. In the localized phase ($V=3$), the system completely loses its spatial diffusion as the potential strength increases further. The fractal dimension approaches zero for all states [c. f. Fig.~\ref{slope.fig}(c)], such as $D_2 = 0.0005$ at $m/L=0.100$ and $D_2 = 0.0010$ at $m/L=0.880$, indicating that the eigenstates become fully localized in the thermodynamic limit.

\setcounter{figure}{0}
\renewcommand{\thefigure}{C\arabic{figure}}
\section{Sorted fractal dimension at $g=0$}
\label{AppNC}

\begin{figure}[htp]
\includegraphics[width=9cm]{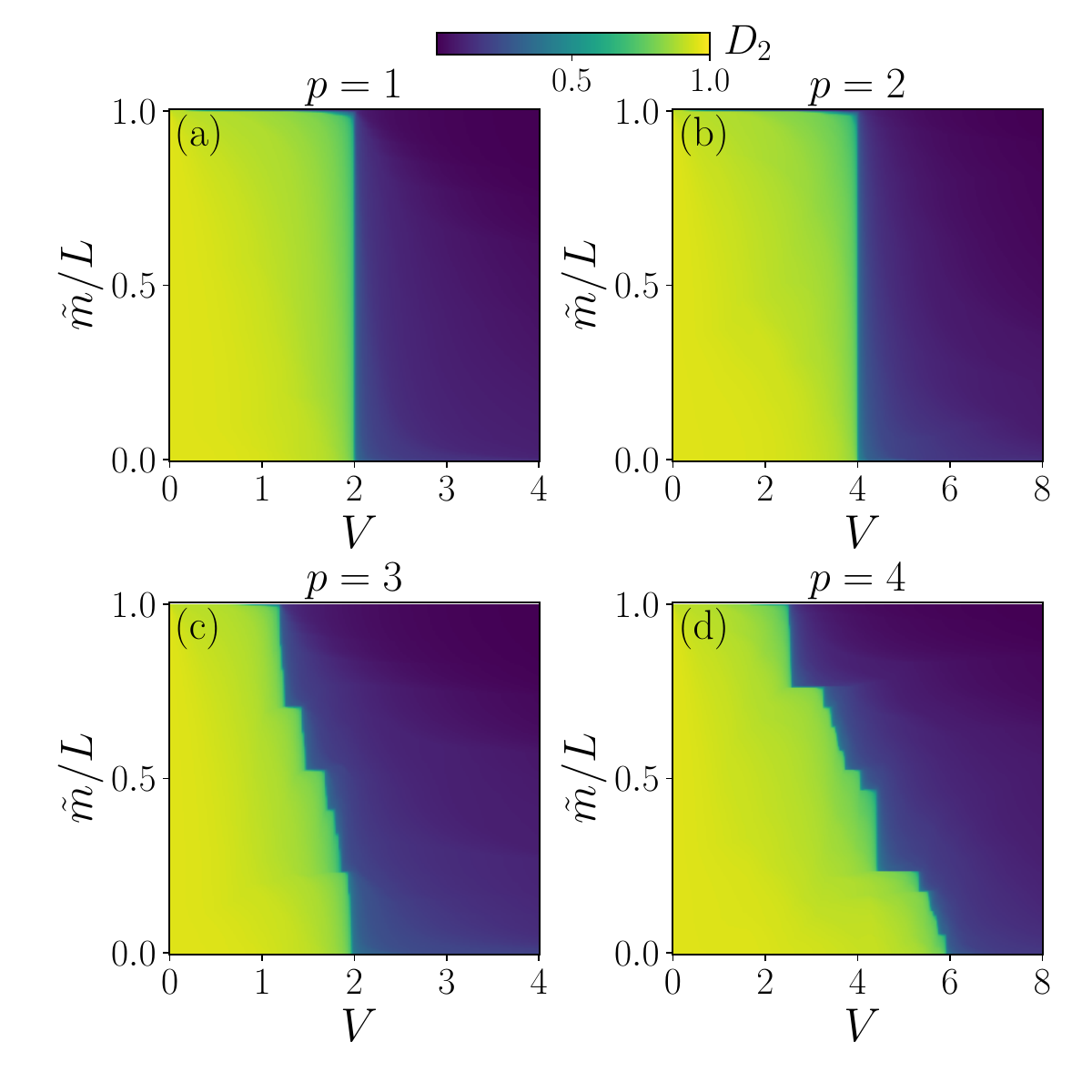}
\centering
\caption{Fractal dimension $D_2$ as a function of the potential strength $V$ in the Hermitian case ($g=0$), with $\beta=(\sqrt{5}-1)/2$ for various power-law exponents. The system size is fixed at $L=4000$. In each subplot, the index $\tilde{m}$ is reordered in ascending order of the IPR to better characterize the localization properties of the eigenstates. Panels (a)-(b), corresponding to $p=1$ and $p=2$, show a sharp and direct transition from an extended phase ($D_2 \approx 1$) to a localized phase ($D_2 \approx 0$). In contrast, panels (c)-(d), for $p = 3$ and $p = 4$, reveal the emergence of a mixed-phase regime, where extended and localized states coexist between the fully extended and fully localized phases. Within the parameter regimes and finite sizes studied here, we do not observe a stable finite interval of fractal states with intermediate $0<D_2<1$.}
\label{IPRD2.fig}
\end{figure}

In this section, we analyze the evolution of the fractal dimension $D_2$ of the eigenstates as a function of the potential strength $V$ in the Hermitian case ($g=0$). To quantitatively characterize the localization transition, we compute $D_2$ for a system size of $L=4000$, as shown in Fig.~\ref{IPRD2.fig}. Although the data are identical to those presented in Fig.~\ref{D2}, the data presentation has been reorganized to more clearly highlight the statistical distribution of eigenstates across different phases. In Fig.~\ref{D2}, the eigenstate index $m$ is ordered by increasing energy. In contrast, in all subplots of Fig.~\ref{IPRD2.fig}, the index $\tilde{m}$ is reordered according to the IPR, from low to high values. This ordering, based on the degree of localization, provides a more intuitive depiction of how the proportions of extended and localized states evolve with increasing potential strength.

The results indicate that the phase transitions of the system exhibit significant differences across various power-law exponents $p$. For $p=1$ and $p=2$, the fractal dimension $D_2$ exhibits an extremely sharp jump with increasing $V$, corresponding to a direct transition from an extended phase ($D_2 \approx 1$) to a localized phase ($D_2 \approx 0$), without a distinct intermediate regime. In contrast, for $p=3$ and $p=4$, a mixed-phase regime emerges. As $V$ increases, the system evolves from an extended phase into a regime where extended and localized states coexist. Within this region, a portion of the eigenstates retains a high fractal dimension, while others exhibit clear localization. With further increase in the potential strength, all states eventually become fully localized. Moreover, numerical analysis shows that no fractal states arise for any power-law exponent, indicating that the phase transitions are governed by the competition and relative fractions of extended and localized states.

\setcounter{figure}{0}
\renewcommand{\thefigure}{D\arabic{figure}}
\section{High IPR at the Silver and Bronze Ratios}
\label{AppA}
To explore how irrational numbers affect the distribution of high inverse participation ratio (IPR) states, we perform a detailed analysis of Hermitian systems under periodic boundary conditions. With the system size fixed at $L=4000$, we consider the silver irrational number ($\beta=\sqrt{2}-1$) and the bronze irrational number ($\beta=(\sqrt{13}-3)/2$) to investigate the phase-transition behavior under odd ($p=3$) and even ($p=4$) power-law potentials.

Figure~\ref{pinfig} presents the phase diagram obtained from the fractal dimension $D_2$ as a function of the potential strength $V$ and the energy levels $m/L$ for the silver [c.f.Fig.~\ref{pinfig}(a) and (b)] and bronze [c.f.~Fig.\ref{pinfig}(c) and (d)] irrational numbers, corresponding to $p=3$ and $p=4$, respectively. The results reveal that all models exhibit three distinct phases: the extended phase, the mixed phase, and the localized phase. Notably, the bronze irrational system displays a richer regime than the silver system, with a more distinct and well-developed mixed phase. This provides a broader parameter space for exploring the properties of intermediate phases.

\begin{figure}[htbp]
\includegraphics[width=18cm]{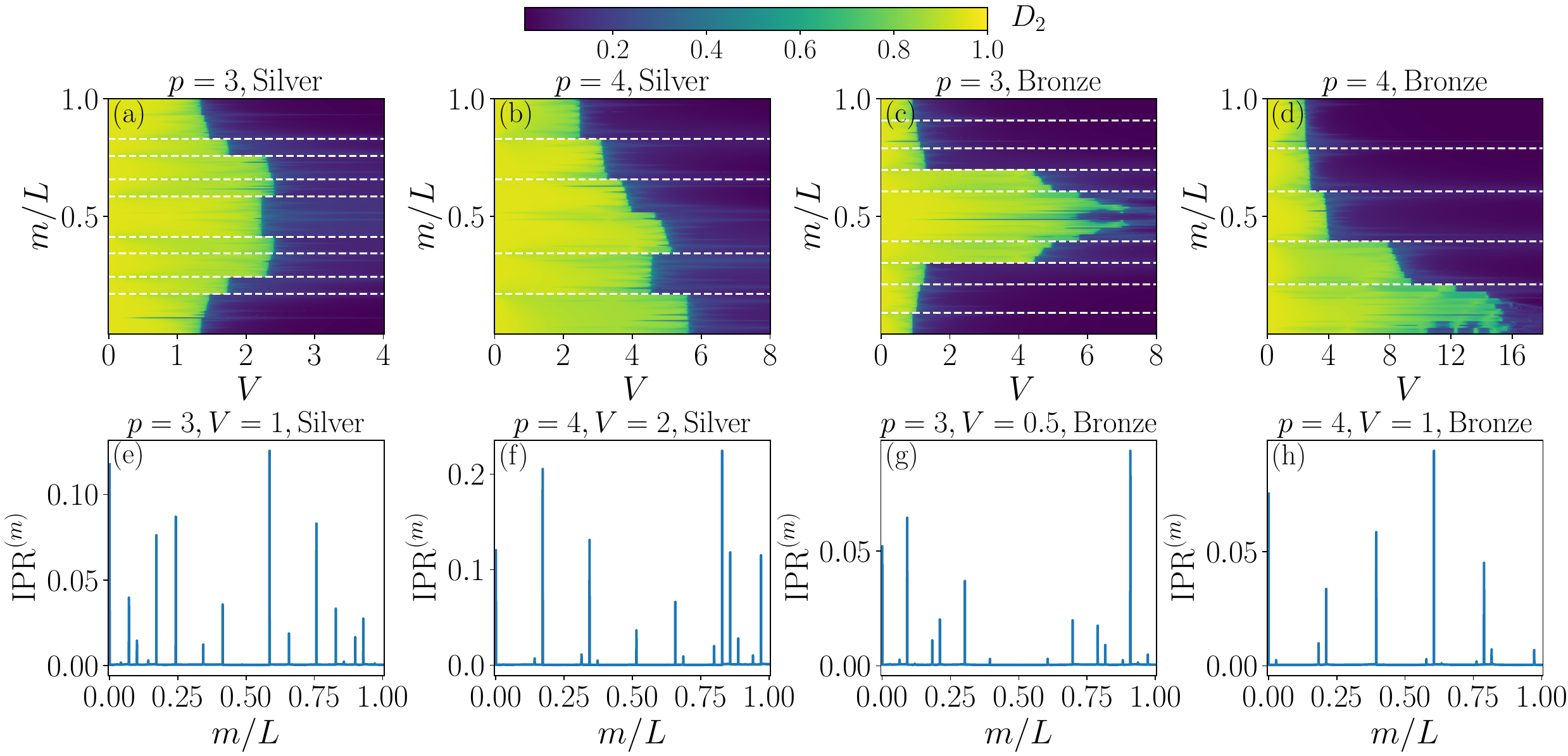}
\centering
\caption{Phase diagrams under periodic boundary conditions with system size $L=4000$ and $g=0$. (a)-(d) Fractal dimension $D_2$ as a function of potential strength $V$ for the silver ratio $(\sqrt{2}-1)$ and the bronze ratio $(\sqrt{13}-3)/2$ at different power-law exponents $p$;   (e)-(h) Distributions of the inverse participation ratio (IPR) of eigenstates at fixed potential strengths $V$ within the extended phase regime.}
\label{pinfig}
\end{figure}

To investigate the distribution of high-IPR states, we conduct a detailed analysis within the extended-phase regime for both the silver and bronze irrational numbers, corresponding to $p=3$ and $p=4$, respectively. We find that the distribution of high-IPR states follows well-defined patterns, as illustrated in the lower panels of Fig.~\ref{pinfig}, for various values of $V$ chosen within the extended phase. These high-IPR states become particularly distinguishable in the non-Fibonacci systems. The positions of these states align closely with the mobility edges, and their distribution patterns precisely follow the analytical expressions given in Eq.~(\ref{na-na}) of the main text. For the $p=3$ case, the index $n$ takes the natural sequence $1,2,3,\ldots$, whereas for the $p=4$ case, $n$ takes only $2,4,6,\ldots$ values. 

Several representative values are listed in Table~\ref{tableA}, corresponding to the white dashed lines in Fig.~\ref{pinfig}. The integer $n$ can be continuously increased until all high-IPR positions are identified. Although different irrational numbers lead to distinct phase boundaries, the underlying mathematical structure determining the high-IPR state positions remains remarkably consistent with Eq.~(\ref{na-na}). 

This finding demonstrates the fundamental role of irrational numbers in quasiperiodic systems from the perspective of microscopic state distributions, establishing a direct connection between irrationality and critical behavior. It provides a robust theoretical basis for understanding phase transitions and the underlying structure of energy spectra. The intrinsic characteristics of irrational numbers are reflected in the distribution patterns of high-IPR states. Accordingly, this result not only confirms the universality of high-IPR state distributions but also highlights the crucial role of irrational numbers in governing phase transitions in quasiperiodic potential systems.

\begin{table}[htbp]
\centering
\caption{Comparison between the numerical and theoretical positions of high-IPR states for the silver \(\beta = \sqrt{2}-1\) and bronze \(\beta = \frac{\sqrt{13}-3}{2}\) irrational numbers at $p=3$ and $p=4$, respectively, with system size $L=4000$.}
\setlength{\tabcolsep}{3pt}
\renewcommand{\arraystretch}{1.1}
\hspace{0 cm}
\makebox[\textwidth][c]{
\begin{minipage}[t]{0.5\textwidth}
\centering
\begin{tabular}{|c|c|c|c|c|}
\hline
\multicolumn{5}{|c|}{$\text{Silver}, p = 3,V=1$} \\
\hline
\multirow{2}{*}{$\hspace{0.1cm}n\hspace{0.4cm}$} & \multicolumn{2}{c|}{$x_n$} & \multicolumn{2}{c|}{$x_n'$} \\
\cline{2-5}
 & Numerical Value & Theory& Numerical Value & Theory\\
\hline
1 & 0.41425 & 0.414214 & 0.586 & 0.585786 \\
\hline
2 & 0.17175 & 0.171573 & 0.8285 & 0.828427 \\
\hline
3 & 0.24275 & 0.242641 & 0.7575 & 0.757359 \\
\hline
4 & 0.34325 & 0.343146 & 0.657 & 0.656854\\
\hline
\multicolumn{5}{|c|}{$\text{Silver}, p = 4,V=2$} \\
\hline
\multirow{2}{*}{$n$} & \multicolumn{2}{c|}{$x_n$} & \multicolumn{2}{c|}{$x_n'$} \\
\cline{2-5}
 & Numerical Value & Theory & Numerical Value & Theory \\
\hline
2 &0.17175 & 0.171573 & 0.8285 & 0.828427 \\
\hline
4 & 0.34325 & 0.343146 & 0.657 & 0.656854 \\
\hline
\end{tabular}
\end{minipage}
\hfill
\begin{minipage}[t]{0.5\textwidth}
\centering
\begin{tabular}{|c|c|c|c|c|}
\hline
\multicolumn{5}{|c|}{$\text{Bronze}, p = 3,V=0.5$} \\
\hline
\multirow{2}{*}{$\hspace{0.1cm}n\hspace{0.4cm}$} & \multicolumn{2}{c|}{$x_n$} & \multicolumn{2}{c|}{$x_n'$} \\
\cline{2-5}
 & Numerical Value & Theory & Numerical Value & Theory \\
\hline
1 & 0.303 & 0.302776 & 0.69725 & 0.697224 \\
\hline
2 &0.3945 & 0.394449 & 0.60575 & 0.605551 \\
\hline
3 & 0.09175 & 0.091673 & 0.9085 & 0.908327 \\
\hline
4 & 0.21125 & 0.211102 & 0.789 & 0.788898\\
\hline
\multicolumn{5}{|c|}{$\text{Bronze}, p = 4,V=1$} \\
\hline
\multirow{2}{*}{$n$} & \multicolumn{2}{c|}{$x_n$} & \multicolumn{2}{c|}{$x_n'$} \\
\cline{2-5}
 & Numerical Value & Theory & Numerical Value & Theory \\
\hline
2 &0.3945 & 0.394449 & 0.60575 & 0.605551 \\
\hline
4 & 0.21125 & 0.211102 & 0.789 & 0.788898 \\
\hline
\end{tabular}
\end{minipage}
}

\label{tableA}

\end{table}

\setcounter{figure}{0}
\renewcommand{\thefigure}{E\arabic{figure}}
\section{Phase transitions in $p=4$ non-Hermitian AAH model}
\label{AppB}
To further verify that the $\mathcal{PT}$ transition in the non-Hermitian power-law AAH model corresponds to the mixed-to-localized phase transition, we investigate the $p=4$ non-Hermitian AAH model with system size $L=987$ using the $\overline{\text{IPR}}$. As shown in Fig.~\ref{p4fig}(a), the system exhibits three distinct phases: extended, mixed, and localized. These phases are further confirmed by the averaged quantities $\overline{\text{IPR}}$, $\overline{\text{NPR}}$, and $\zeta$ at $g=0.3$, presented in Fig.~\ref{p4fig}(b). In the regime $3.5 < V < 7.7$, all three quantities remain finite, signifying the coexistence of extended and localized states that define the mixed phase.
The nature of the three phases and their correspondence to the complex energy spectrum are illustrated in Fig.~\ref{p4fig}(c) and (d). The $\mathcal{PT}$ transition, marked by the black dashed line at $V=7.7$ coincides with the boundary between the mixed and localized phases. As shown in Fig.~\ref{p4fig}(c), the energy levels in the extended phase are predominantly imaginary, while in the mixed phase they become partially complex, displaying a sharp discontinuity at the transition point $V=3.5$ (denoted by the red dashed line). These results are fully consistent with the behavior presented in Fig.~\ref{p4fig}(b).

\begin{figure}[tp]
\centering
\includegraphics[width=11.3cm]{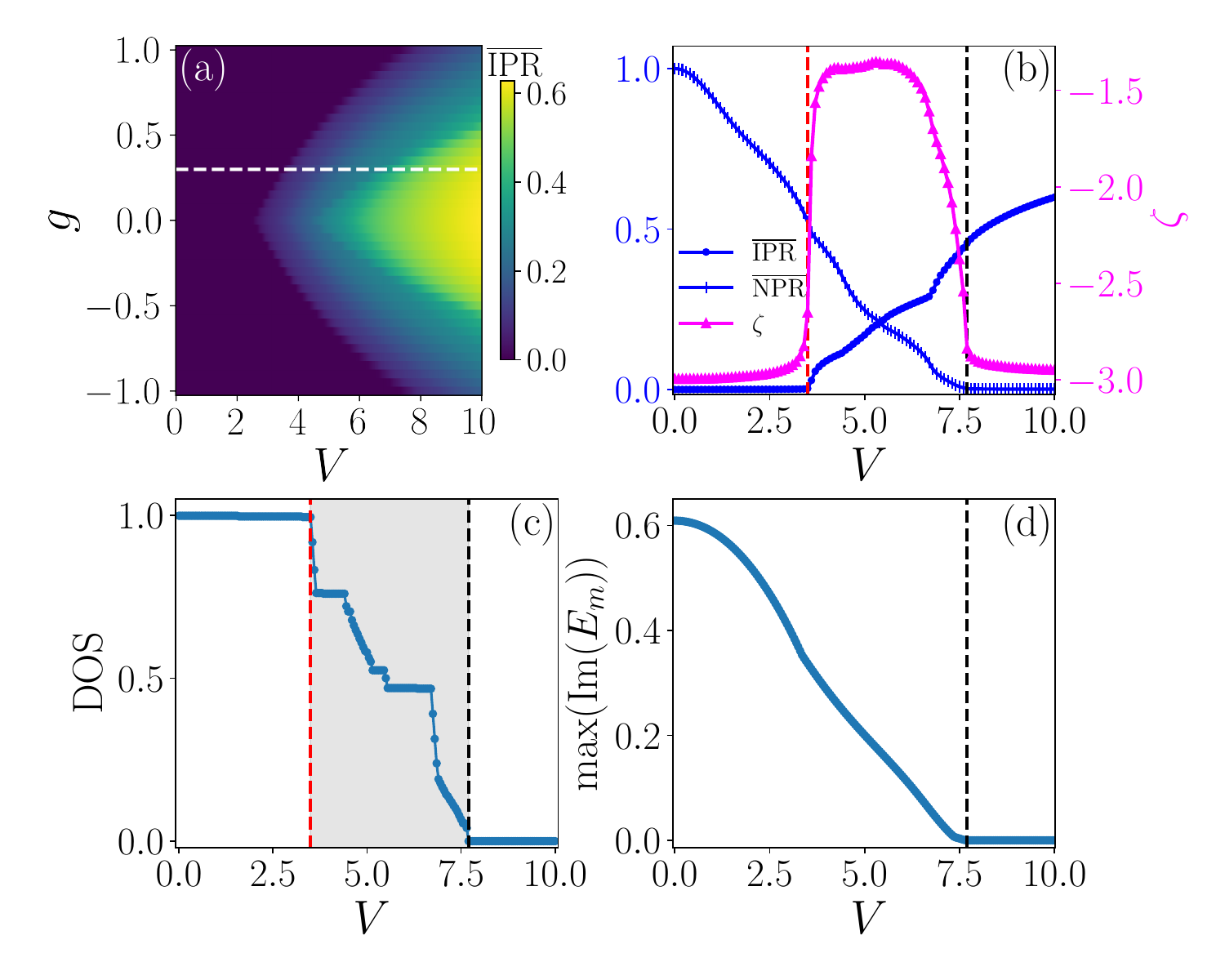} \centering
\caption{Phase transitions of the $p=4$ non-Hermitian AAH model. (a) Phase diagram as a function of $V$ at $L = 987$. The white dashed line correspond to $g=0.3$. (b) $\overline{\text{IPR}}$ and $\overline{\text{NPR}}$ (blue curves) and $\zeta$ (pink curve), as a function of $V$ at $g=0.3$; (c) Density of states (DOS) versus $V$ at $g=0.3$; (d) Maximum imaginary part of the eigenvalues $\max(\operatorname{Im}(E_\text{m}))$ with respect to $V$ at $g=0.3$. The red and black dashed lines indicate the extended-to-mixed state transition and the \(\mathcal{PT}\) transition at $V = 3.5$ and $V = 7.7$, respectively.}
\label{p4fig}
\end{figure}

\setcounter{figure}{0}
\renewcommand{\thefigure}{F\arabic{figure}}
\section{Sorted fractal dimension at $g=0.3$}
\label{AppND}

\begin{figure}[htp]
\includegraphics[width=9.6cm]{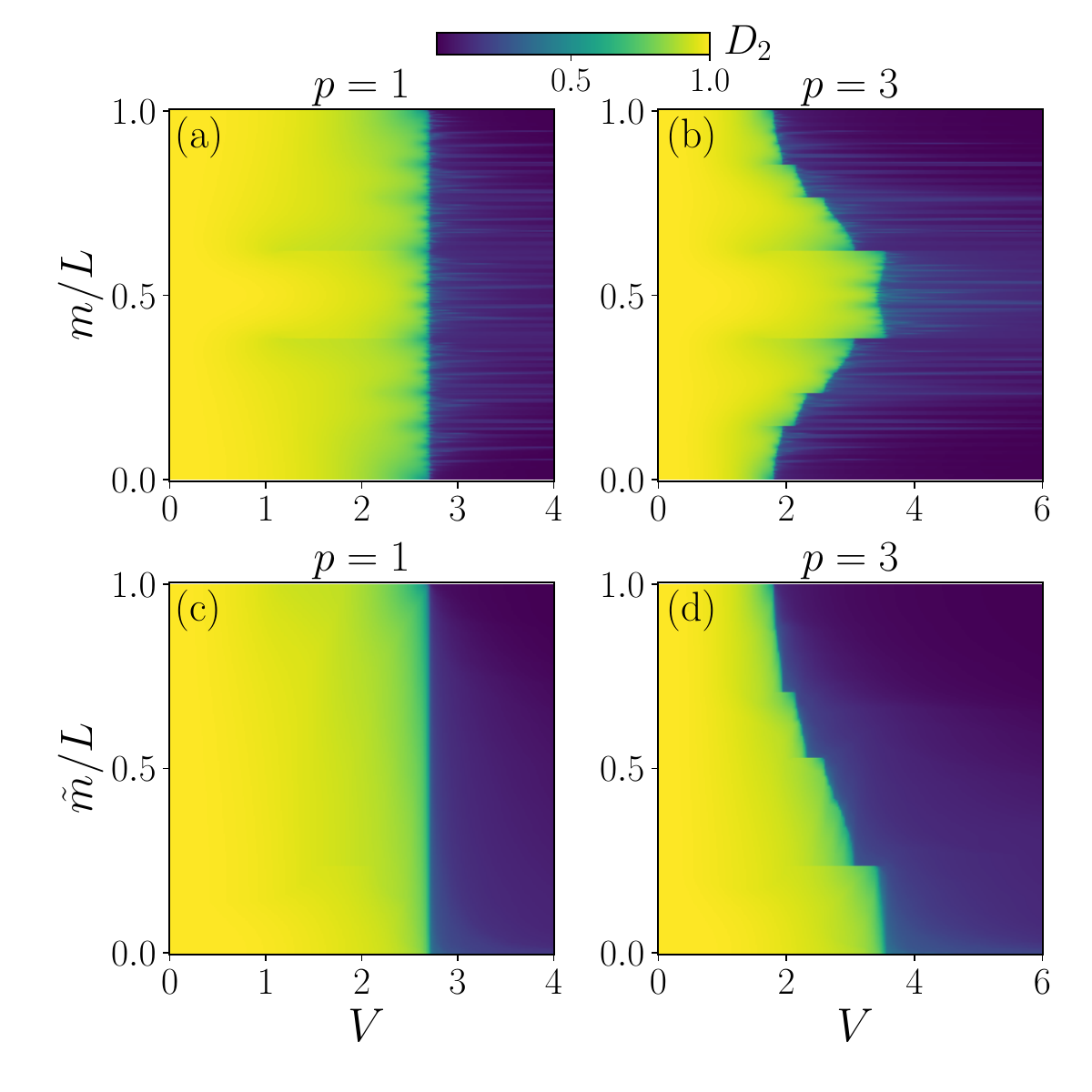}
\centering
\caption{Fractal dimension $D_2$ as a function of the potential strength $V$ in the non-Hermitian case ($g=0.3$), at $\beta=(\sqrt{5}-1)/2$ for different power-law exponents. The system size is fixed at the Fibonacci number $L=987$. Panels (a) and (c) correspond to $p=1$ under two different sorting schemes, while panels (b) and (d) correspond to $p=3$. In the upper panels (a)-(b), the energy-level index $m$ follows the ascending order of the real parts of the eigenvalues. In the lower panels (c)-(d), the reordered index $\tilde{m}$ is sorted in ascending order of the inverse participation ratio (IPR). The results show that for $p=1$, the system undergoes a direct transition from an extended phase to a localized phase. In contrast, for $p=3$, a distinct mixed-phase regime emerges, characterized by the coexistence of extended and localized states before the system becomes fully localized.}
\label{gD2.fig}
\end{figure}

In this section, we compute the fractal dimension $D_2$ of the eigenstates as a function of the potential strength $V$ to investigate the localization properties of the system in the non-Hermitian case ($g=0.3$), with the system size fixed at the Fibonacci number $L=987$. To provide a more comprehensive characterization of the evolution of the eigenstates, we employ two complementary indexing schemes. In panels (a) and (b) of Fig.~\ref{gD2.fig}, the energy-level index $m$ follows the ordering of the eigenenergies, whereas in panels (c) and (d), the reordered index $\tilde{m}$ is sorted according to the IPR. Although the data are identical, the presentation has been reorganized to more clearly highlight the statistical distribution of eigenstates across different phases, enabling a more sensitive characterization of their localization and delocalization properties.

As illustrated in Fig.~\ref{gD2.fig}, panels (a) and (c) present the numerical results for the power-law exponent $p=1$, while panels (b) and (d) correspond to $p=3$. Combined with the previously discussed results in Fig.~\ref{MIPRfig}, it is evident that mixed states persist even in the presence of non-Hermiticity. Specifically, for $p=1$, the system undergoes a direct transition, with the eigenstates exhibiting an abrupt change from a fully extended phase to a globally localized phase as the potential strength $V$ increases. In contrast, for $p=3$, the system exhibits a distinct mixed-phase regime, in which extended and localized states coexist before the system enters a fully localized phase. This feature in the distribution of eigenstates is also clearly captured in the $\tilde{m}$-sorted fractal dimension.


\end{widetext}

\end{document}